\documentclass[10pt, twoside]{article} 

\usepackage[numbers]{natbib}
\usepackage[affil-it]{authblk}

\usepackage[T1]{fontenc} 
\usepackage{microtype} 

\usepackage[columnsep=25pt]{geometry}
\usepackage{multicol} 
\usepackage[hang, small,labelfont=bf,up,textfont=it,up]{caption} 
\usepackage{booktabs} 
\usepackage{float} 

\usepackage{abstract} 

\usepackage{titlesec} 
\renewcommand\thesection{\Roman{section}} 
\renewcommand\thesubsection{\Roman{subsection}} 
\titleformat{\section}[block]{\large\scshape\centering}{\thesection.}{1em}{} 
\titleformat{\subsection}[block]{\large}{\thesubsection.}{1em}{} 

\usepackage{fancyhdr} 
\usepackage{amsmath}
\usepackage{amsthm}
\usepackage[english]{babel}
\usepackage{graphicx}
\usepackage{anysize}
\usepackage{amsfonts} 
\usepackage{amssymb} 
\usepackage{float}
\usepackage{times}
\usepackage{mathtools}
\usepackage{color}
\usepackage{mathrsfs}

\usepackage{hyperref} 

\newcommand{\deriv}[1]{\frac{d}{d #1}}
\newcommand{\DERIV}[2]{\frac{d #1}{d #2}}

\newcommand{\PD}[2]{\frac{\partial #1}{\partial #2}}

\title{Cosmological history in York time: inflation and perturbations}
\author{\large Philipp Roser\thanks{proser@clemson.edu}\, and Antony Valentini\thanks{antonyv@clemson.edu}}
\affil{\small Department of Physics and Astronomy,\\ Clemson University, Kinard Laboratory,\\ Clemson, SC 29631-0978, USA}
\date{\vspace{-0.5cm}}		

\begin{document}
\maketitle
\begin{abstract}
The constant mean extrinsic curvature on a spacelike slice may constitute a physically preferred time coordinate, `York time'. One line of enquiry to probe this idea is to understand processes in our cosmological history in terms of York time. Following a review of the theoretical motivations, we focus on slow-roll inflation and the freezing and Hubble re-entry of cosmological perturbations. We show how the mathematical account of these processes is distinct from the conventional account in terms of standard cosmological or conformal time. We also consider the cosmological York-timeline more broadly and contrast it with the conventional cosmological timeline.
\end{abstract}

\begin{multicols}{2}

\section{Introduction}\label{sec.introduction}

At the classical level general relativity adequately accounts for empirical observations of gravitational phenomena. However, attempts to take the theory into the quantum realm are thwarted by a series of conceptual and technical difficulties jointly referred to as the `problem(s) of time' \citep{Isham1992, Kuchar2011, Anderson2012}. These difficulties are, roughly speaking, the result of the desire to retain refoliation invariance (the freedom to `slice' spacetime into spatial surfaces of constant time in any desirable manner) and reparameterisation invariance (the freedom to label spatial slices by a monotonic but otherwise arbitrary `time' parameter) in one form or another at the quantum level. Demanding that the quantum theory be independent of the foliation chosen prior to quantisation presents a difficulty for conventional quantisation methods.

One way to overcome these issues is to give up on refoliation and reparameterisation invariance by proposing the existence of a physically preferred spacetime foliation and time parameter. Of course, the theory of quantum gravity resulting from the quantisation of such a physical-time formalism must in the appropriate classical limit recover the full set of symmetries of the classical theory --- classically no hints of a preferred time parameter have been observed. Even if this is ultimately possible, adopting a preferred time parameter may seem like a high theoretical price to pay. But in any case empirical evidence will be the ultimate arbiter.

A fundamental time-space split may be unavoidable, not merely for technical reasons relating to the problem of time but also for concrete physical reasons. Quantum physics is in some sense non-local (the precise physical meaning of which is still controversial). It has been argued, for example, that a description of quantum entanglement requires a preferred notion of simultaneity. Whatever the final form of the theory of quantum gravity, it is likely to include interaction between geometric elements and matter. The quantum theory of the latter is relatively well understood and involves the possibility of such entanglement. 

There are hints even from non-gravitational high-energy physics that local Lorentz invariance 
may not be fundamental \citep{Liberati2013, Valentini2008inCraigSmith, Afshordi2015} and testing possible violations (such as not Lorentz-invariant standard-model extensions) is an ongoing enterprise \citep{ColladayKostelecky1997,Kostelecky2004,Amelino-Camelia2004,LiberatiMaccione2009}. For example, quantum field theories are usually considered only applicable up to some cut-off energy, which avoids divergences. However, the notion of an energy cut-off is not Lorentz invariant. While the true microscopic structure that approximates a continuous spacetime (with the symmetries of general relativity in the classical limit) may well be more exotic than a physically fundamental time-space split, the latter is not implausible and does adequately address these issues from field theory. Recent developments in the area of gravitational theories with a preferred foliation include, for example, Ho\v rava-Lifshitz gravity \citep{Horava2009,Visser2011}, whose preferred foliation in its low-energy limit has been argued to match the `York-time' foliation explored here \citep{Afshordi2009}. 

For a number of theoretical reasons reviewed in section \ref{sec.Yorktime} a foliation of spacetime into spatial slices of constant extrinsic scalar curvature (or `constant mean curvature'), parameterised by `York time' --- effectively the rate of local spatial contraction with respect to proper time --- is a uniquely viable candidate. The primary reason is related to the role of constant-mean-curvature slicing and York time in the initial-value problem of general relativity. In addition, such a foliation has a variety of desirable theoretical properties.

The York-time gravitational theory is classically equivalent to general relativity\footnote{Not all solutions of general relativity permit a global York-time slicing. Therefore, the proposal that York time be the physically fundamental time parameter rules out certain solutions of general relativity. However, for the most part these are arguably physically pathological, involving for example closed time-like lines \citep{MarsdenTipler1980}.} and may be derived via Hamiltonian reduction \citep{ChoquetBruhatYork1980} starting from the Arnowitt-Deser-Misner (ADM) formalism \citep{ADM1962, MisnerThorneWheeler1973}, though this is difficult in full generality since it involves solving a complicated elliptic partial differential equation. For a homogeneous and isotropic minisuperspace model the reduction was undertaken in ref.\ \citep{RoserValentini2014a} by the present authors and a corresponding analysis for an anisotropic model was performed by one of us in ref.\ \citep{Roser2015a}. In order to make connection with our actual universe, the project of developing cosmological perturbation theory via a perturbative approach to Hamiltonian reduction was begun in ref.\ \citep{Roser2015b}.

While such formal developments are required in order to ultimately obtain a self-contained theory, it can be insightful to gain an understanding from the York-time perspective of phenomena that are already well understood in the conventional description, such as processes in the cosmological history of our universe. This is the purpose of the present paper. Following a review of York time in section \ref{sec.Yorktime}, we focus on a York-time description of inflation (section \ref{sec.inflation}), the `freezing' of cosmological perturbations during this period (section \ref{sec.modefreezing}) and finally their Hubble `re-entry' during the radiation-dominated era (section \ref{sec.raddomuniverse}). We provide some numerical estimates for York-time scales in section \ref{sec.timeline} before presenting our conclusions in the final section (\ref{sec.conclusion}). Of course, since our formalism is classically equivalent to standard general relativity we do not find any truly new phenomena. Instead we illustrate how the same, well-understood phenomena arise via distinct mechanisms.


\section{York time}\label{sec.Yorktime}

In order to implement the time-space split of spacetime and obtain an explicit Hamiltonian generating dynamics with respect to the chosen time parameter, one would begin with the Hamiltonian constraint obtained via the ADM `3+1' description \citep{ADM1962} and solve it for the momentum conjugate to the chosen intrinsic or extrinsic variable that is to take the role of time. This conjugate momentum would then be the (negative of) the physical Hamiltonian density. Such a procedure is known as Hamiltonian reduction. The value of York time $T$ is equal to the scalar extrinsic curvature $K$ up to a constant numerical factor chosen such that the conjugate momentum $P_T$ of $T$ is exactly the negative of the local volume element,
\begin{equation} T\equiv \frac{2g_{ij}\pi^{ij}}{3\sqrt{g}},\qquad P_T\equiv -\sqrt{g}, \end{equation}
where $g_{ij}$ is the spatial 3-metric, $g$ its determinant and $\pi^{ij}$ its conjugate momentum as obtained in the ADM description.\footnote{The relation between the momenta and the extrinsic curvature is given by  $\pi^{ab}=\sqrt{g}(Kg^{ab}-K^{ab})$.}

More intuitively, up to a numerical constant York time $T$ is equal to the fractional rate of contraction with respect to proper time (normal to the hypersurface) of a local 3-space volume element. This parameter is constant across a spatial slice (a prerequisite for interpreting the slice as `space at some given time') if the extrinsic scalar curvature is constant across the slice. In the context of cosmology York time is the Hubble parameter $H$ up to a negative constant,
\begin{equation} T = -(4\pi G)^{-1} H = -2M_{Pl}^{2} H, \label{eq.THincosm}\end{equation}
where $G$ is the gravitational constant and $M_{Pl}$ the reduced Planck mass. Throughout this paper we use units such that $c=1$ and $\hbar=1$. The monotonicity of $T$, a requirement for it to function as a time parameter at all, is guaranteed if the equation of state parameter $w$ obeys $w>-1$. This is a result of substituting the mass-energy conservation equation with energy density $\rho$ and pressure $p$,
\begin{equation} \dot{\rho} = -3H(\rho+p), \end{equation}
into the time derivative of the first Friedmann equation,
\begin{equation} 2H\dot{H} = (3M_{Pl}^2)^{-1}\dot{\rho}.\end{equation}
A dot denotes a derivative with respect to $t$. All conventional matter content obeys $w>-1$. For example, in the case of a universe dominated by a scalar field one has 
\begin{equation}\rho=\tfrac12\dot{\phi}^2+V(\phi),\qquad p=\tfrac12\dot{\phi}^2-V(\phi),\end{equation}
which implies $w\geq-1$ with equality only in the (arguably unphysical) limiting case when the kinetic contribution vanishes exactly, a scenario corresponding to a first approximation to slow-roll inflation. This we will study in detail in section \ref{sec.inflation}.

A major motivation for considering that the constant-mean-curvature foliation (parameterised by $T$) may be physically fundamental is its role in the initial-value problem of general relativity \citep{York1972, ChoquetBruhatYork1980}. The initial-value problem is the task to specify a complete set of Cauchy data (consisting of a specification of $g_{ij}$ and $\pi^{ij}$ at each spatial point) on a spatial slice such that the Hamiltonian and momentum constraints are satisfied and the evolution of the 3-geometry is uniquely determined. In the early 1970s York showed, building on earlier work of Lichnerowicz and others, that on a slice of constant scalar extrinsic curvature such a set of data may be obtained by first specifying $g_{ij}$ and $\pi^{ij}$ such that they satisfy the momentum constraints only (which is relatively straightforward) and then performing a conformal transformation $g_{ij}\rightarrow \bar{g}_{ij}=\chi^4g_{ij}$, $\sigma^{ij}\rightarrow \bar{\sigma}^{ij}=\chi^{-4}\sigma^{ij}$, with $\sigma^{ij}$ denoting the trace-free part of $\pi^{ij}$, in which the scale function $\chi(x)$ is determined by the requirement that the rescaled variables satisfy the Hamiltonian constraint also. The resulting Cauchy data $\{\bar{g}_{ij},\bar{\pi}^{ij}\}$ is independent of the initial choice of scale. That is, had we started with other variables $g^\prime_{ij}$, $\pi^{\prime ij}$ related to our original choice by another conformal transformation, $g^\prime_{ij}=\omega^4 g_{ij}$, $\pi^{\prime ij} = \omega^{-4}\pi^{ij}$, the resulting data $\{\bar{g}_{ij},\bar{\pi}^{ij}\}$ would have been the same.

This result suggests that spatial scale does not have the same dynamical status as the remaining, `shape' variables representing the local geometry. Instead scale and its conjugate momentum, following a change of variables to $T$ and $P_T$, take the role of time and Hamiltonian density. Taking this idea seriously in conjunction with the requirement that the dynamics be formulated entirely in terms of dimensionless quantities has recently led to the development of an intrinsically `3+1' theory of gravity called Shape Dynamics \citep{GomesGrybKoslowski2011, BarbourKoslowskiMercati2013ProbOfTime, Mercati2014}, which bears a close relationship to general relativity in the York-time description.

A York-time foliation also has other properties that make it a plausible candidate on which to build a fundamental theory of quantum gravity. For example, in the presence of black holes the foliation `wraps around' the singularities \citep{QadirWheeler1985}, so that all singularities are part of one final singular surface at $T=\infty$ (coinciding with the `big crunch').\footnote{In a forever expanding scenario, such as spatially flat cosmologies with conventional matter content (where the equation of state parameter $w>-1$) York time remains negative throughout cosmological history. Values $T>0$ correspond to a possible symmetric extension of this history. In fact, such an extension may be necessary for consistent quantisation \citep{Roser2015CosmExtension} given the behaviour of the constant-mean-curvature foliation around singularities.} Hence the geometrical degrees of freedom on any slice corresponding to a finite value of $T$ are always non-singular. Because singularities are avoided in the classical theory there is no open question about how such singularities might arise in the appropriate classical limit of a corresponding quantum theory. Further motivation from quantum theory was discussed in ref.\ \citep{Valentini1996}, where it was argued that constant-mean-curvature slicing may function as a fundamental slicing for certain kinds of hidden-variable theories and their intrinsically non-local entities. 

We conclude that if quantum gravity can indeed be obtained via the quantisation of a 3+1 theory of gravity derived from general relativity, then York time might provide the appropriate foliation and parameterisation. Before undertaking such an ambitious endeavour, however, it may be useful to understand basic cosmological processes in terms of York time. This will be our concern for the rest of this paper.


\section{Inflation} \label{sec.inflation}

The theory of inflation poses an immediate puzzle for a description in terms of York time. During an exact de~Sitter phase the spatial expansion of the universe is exponential, corresponding to a constant Hubble parameter and therefore constant York time. That is, the universe appears to expand by a finite amount while no time passes. The puzzle is resolved immediately if we recall that in a realistic inflationary phase the Hubble parameter is constant only to a first approximation and is, in fact, decreasing, albeit only by a small amount. The implication is that an inflationary period corresponds to a very short time interval and furthermore that the York-time description must use a higher order of approximation than is conventionally used in the standard picture.

Conventional slow-roll inflation assumes that during this period the matter content of the universe is dominated by a scalar field, whose contribution to the energy density $\rho = \frac12\dot{\phi}^2+V(\phi)$ is in turn dominated by the field's potential function $V(\phi)$, where $\dot{\phi}$ denotes the derivative of $\phi$ with respect to conventional cosmological time $t$. Specifically, one assumes the `slow-roll conditions'
\begin{equation} \tfrac12\dot{\phi}^2\ll V(\phi),\qquad |\ddot{\phi}|\ll\left|\PD{V}{\phi}\right|, \end{equation}
while the dynamics is determined by the Friedmann and Klein-Gordon equations,
\begin{align}
 3H^2 &= M_{Pl}^{-2} \rho \label{eq:Friedmann}\\
 \ddot{\phi}+3H\dot{\phi}+\PD{V}{\phi} &= 0. \label{eq:KleinGordon}
\end{align}
Provided the slow-roll conditions hold, in terms of York time $T=-2M_{Pl}^2H$ the Friedmann equation \ref{eq:Friedmann} becomes
\begin{equation}\tfrac34 M_{Pl}^{-2} T^2\approx V(\phi), \end{equation}
which suggests that the numerical value of the potential effectively provides a measure of York time. Approximating the potential to be roughly constant is therefore not a viable description. 

Instead one proceeds to calculate the next-order, time-dependent correction to $H$, and hence $T$. At `zeroth' order one ignores the contribution of the kinetic term to $\rho$, so that equation \ref{eq:Friedmann} becomes 
\begin{equation} H \approx \sqrt{V(\phi)/3M_{Pl}^2}, \label{eq:FriedmannZerothOrder}\end{equation}
which can be substituted into the Klein-Gordon equation \ref{eq:KleinGordon}. Applying the second slow-roll condition, the second-derivative term $\ddot{\phi}$ can be neglected, so one is left, in general, with a non-linear first order differential equation, which can be solved in many relevant cases to yield a first time-dependent approximation $\phi_1(t)$ for the scalar field. For example, for $V(\phi)=\lambda\phi^b$, the solution is 
\begin{equation} \phi_1(t) = \left[\left(\frac{b}{2}-2\right)\left(b\sqrt{{\lambda M_{Pl}^2}/{3}}\,t+A\right)\right]^{1/(2-\frac{b}{2})}, \label{eq:phisolutionpolynomial} \end{equation}
with $A$ being an integration constant that must obey certain conditions that follow from the slow-roll conditions, namely that the constant contribution to the potential $V(\phi_1)$ is large compared to the time-dependent one (we will show this explicitly below for a particular example). This solution $\phi_1$ can then be subsituted back into the Friedmann equation \ref{eq:Friedmann} to give the next approximation $H_1(t)$ to the Hubble parameter and therefore to the York parameter.\footnote{One could repeat the procedure iteratively but in general this is not necessary for an approximate description of the homogeneous cosmological background and would in any case not yield higher degrees of accuracy due to our neglect of the contribution of other matter.} The function $T(t)$ obtained can now be inverted to find $\phi$ in terms of $T$ and give a `York-time only' account of the evolution during inflation.

To illustrate this, consider a particular example of an inflationary potential, namely a case of so-called large-field inflation, $V(\phi)=\frac12 m^2\phi^2$. This simple choice helps to illustrate the method of obtaining a York-time description since unlike many other cases it leads to straightforward algebra (although it is not a choice of potential favoured by current observation \citep{Planck2015_Overview}). One finds,
\begin{equation} \phi_1(t) = \mp\sqrt{\tfrac43M_{Pl}^2m^2}\,t + \phi_0, \qquad\text{for }\phi_0\gtrless0 \label{eq:phi1solution}\end{equation}
where $\phi_0$ is the value of $\phi$ at $t=0$, which must be chosen such that the approximation is valid at this time. We will justify the condition on the sign, which arises due to the appearance of `$\sqrt{\phi^2}$' in the Klein-Gordon equation, shortly. The first slow-roll condition then says
\begin{equation} \tfrac23 M_{Pl}^2m^2\ll m^2\left(\phi_0+\sqrt{\tfrac43M_{Pl}^2m^2}\,t\right)^2, \label{eq:1stSRCinExample}\end{equation}
which must be satisfied throughout inflation, in particular at $t=0$, that is,
\begin{equation}\phi_0^2\gg \tfrac23 M_{Pl}^2. \label{eq:t0SRC}\end{equation}
The constant contribution $\phi_0$ dominates the scalar field until $t\approx(\frac43M_{Pl}^2m^2)^{-\frac12}$, by which point the approximation has ceased to be valid, presumably corresponding to a time well after the end of inflation. While the approximation remains valid the sign of $\phi$ is just the sign of $\phi_0$, hence the sign dependence in equation \ref{eq:phi1solution}.

In this example the correction to $\rho$ due to the kinetic energy contribution is constant, 
\begin{equation} \frac{1}{3M_{Pl}^2}\cdot\tfrac12\dot{\phi}^2 = \tfrac29 m^2, \label{eq:KEcontribution}\end{equation}
while time dependence arises via the potential term,
\begin{equation} V(\phi_1(t)) = m^2\phi_1^2 = m^2\left(\phi_0+\sqrt{\tfrac43M_{Pl}^2m^2}\,t\right)^2.\end{equation}
This is consistent with the approximation since the time-independent contribution dominates by virtue of equation \ref{eq:1stSRCinExample}. Substituting these back into the Friedmann equation gives
\begin{equation} H_1^2 = H_0^2\left[\left(1\mp\phi_0^{-1}\sqrt{\tfrac43 M_{Pl}^2m^2}\right)^2 + H_1^{-2}\cdot\tfrac29m^2\right],\end{equation}
where
\begin{equation} H_0^2 = \frac{1}{3M_{Pl}^2}m^2\phi_0^2 \end{equation}
is the initial, constant zeroth-order approximation to the Hubble parameter as given by equation \ref{eq:FriedmannZerothOrder}. From this one obtains the expression for York time as a function of cosmological time, 
\begin{equation} T(t) = -2M_{Pl}^2\cdot H_0\left(1+\frac{m^2}{9H_0^2}-\sqrt{\tfrac43 M_{PL}^2m^2}\frac{t}{|\phi_0|}\right).\end{equation}
This expression may be straightforwardly inverted in order to express $\phi_1$ as a function of $T$ rather than $t$,
\begin{equation} t = \left(\tfrac43 M_{Pl}^2m^2\right)^{-\frac12}|\phi_0|\left(1+\frac{m^2}{9H_0^2}-\frac{|T|}{2M_{Pl}^2H_o}\right). \end{equation}

The primary characteristic of inflation is the rapid (approximately exponential) growth of the background scale factor $a$. In the York-time formalism $a$ ceases to be a dynamical variable however since the `volume' $a^3$ is the momentum conjugate to the time parameter (up to a sign). The volume, up to some suitable normalisation, is therefore given by the numerical value of the reduced York-time Hamiltonian, which for a spatially flat cosmology\footnote{For a closed cosmology an analogous (although algebraically more complicated) expression can be derived \citep{RoserValentini2014a}.} is given by (see ref.\ \citep{RoserValentini2014a}) 
\begin{equation} a^3 = \mathcal{H}_T = \sqrt{{\tfrac12p_{\phi}^2}\big[{3T^2/(4M_{Pl}^2)-V(\phi)}\big]^{-1}}, \end{equation}
where $p_\phi$ denotes the momentum conjugate to the scalar field. The York-time derivative $a^\prime = da/dT$ of the scale factor is then easy to obtain (since $d\mathcal{H}_T/dT=\partial\mathcal{H}_{T}/\partial T$):
\begin{equation} \frac{a^\prime}{a} = -\frac{\frac34M_{Pl}^{-2} T}{\frac34M_{Pl}^{-2} T^2-V(\phi)}. \label{eq:Yubble} \end{equation}
The expression in the denominator is nothing but the total kinetic energy as is apparent from the Friedmann equation (\ref{eq:Friedmann}). In the example above it is constant and so one can easily calculate the number of e-folds that correspond to some inflationary interval $(T_i,T_f)$,
\begin{equation} N_e = \int_{T_i}^{T_f}dT\;\frac{a^\prime}{a} = \tfrac12 \kappa \left(T_i^2-T_f^2\right), \label{eq:efolds}\end{equation}
where
\begin{equation} \kappa \equiv \frac{\frac34M_{Pl}^{-2}}{\frac34M_{Pl}^{-2} T^2-V(\phi)} = \text{ const.} \label{eq:kappadef}\end{equation}
has been defined for convenience. For other choices of potential the solution $\phi_1(t)$ is in general not a linear function of $t$ and so the kinetic energy is not constant. However, even then one can take the solution \ref{eq:phisolutionpolynomial} and expand it, dropping terms of second and higher order in $t$ since these are small compared to the constant and linear terms by the first slow-roll condition. As a result the non-constant contribution to the kinetic energy is negligible even compared to the already small constant contribution. Equation \ref{eq:efolds} holds, therefore, more generally. 

Equation \ref{eq:Yubble} is then trivially solved, giving
\begin{equation} a(T) = a_0 e^{-\frac{\kappa}{2}T^2}, \label{eq:aduringinflation}\end{equation}
where the integration constant $a_0$ hypothetically denotes a normalisation scale corresponding to the value of $a$ in the infinite $t$-future ($T=0$), although such an interpretation is not appropriate since the inflationary description only holds for a limited duration.


\section{Freezing of perturbations during inflation}\label{sec.modefreezing}

Inflation originated as an explanation for various puzzles relating to cosmological initial conditions such as the horizon and flatness problems \citep{Guth1981}. Since then, however, its greatest success has arguably been to explain the near scale-invariant power spectrum of cosmological perturbations. Observations of the cosmic microwave background continue to narrow down the range of possible inflaton potentials \citep[for example, ref.\ ][]{Planck2015_Overview}.\footnote{Despite its broad appeal,  there are rival theories to inflation aiming to explain our cosmological observations, such as, for example, bouncing cosmologies \citep[see][for a recent review]{BrandenbergerPeter2016}.} Cosmological perturbation theory is one of the cornerstones of modern cosmology, accounting for both structure formation and CMB inhomogeneities. Understanding the evolution of perturbations in terms of York time is therefore essential if $T$ is to be considered the physically fundamental time parameter.

The development of formal cosmological perturbation theory in the York-time reduced-Hamiltonian description was begun in ref.\ \citep{Roser2015b} and lies outside the scope of this paper. More immediate insights may be gained by `translating' the standard equations governing cosmological perturbations into York time via the appropriate change of variable $t\rightarrow T(t)$. The physical content of the theory remains unchanged. Nonetheless the mechanisms for certain processes, such as the `freezing' of (Fourier modes of) perturbations, are rather different.

While a more detailed analysis would lead us to consider scalar and tensor perturbations separately, for our present purposes it suffices to consider the single sufficiently general equation of motion for a Fourier mode with wave number $k$,
\begin{equation} \frac{d^2y_k}{d\eta^2} + \left(k^2-\frac{1}{a}\frac{d^2a}{d\eta^2}\right)y_k = 0, \label{eq:ModeEqInEta}\end{equation}
where $\eta$ denotes conformal time. The variable $y_k$ here might refer to a particular tensor mode after rescaling by the scale factor. That is, we may parameterise the two independent degrees of freedom of the tensor perturbation as
\begin{equation} \delta g_{ij} = a^2\left( h_1 e_{ij}^1 + h_2 e_{ij}^2 \right),\end{equation}
where $\{e^1_{ij},e^2_{ij}\}$ form an orthonormal basis of the tensor perturbations of the spatial metric $g_{ij}$, then rescale to $y=\frac14M_{Pl}^2 ah_1$ (and similar for $h_2$) and perform a Fourier transform to arrive at equation \ref{eq:ModeEqInEta} (the equation of motion derived from the Einstein-Hilbert action). A similar equation arises in the case of scalar perturbations (scalar metric and scalar-field perturbations, for example) when expressed in appropriate variables (rescaling the field $\phi$ by the scale factor $a$) and with an appropriate gauge choice \citep{MukhanovFeldmanBrandenberger1992}. Equation \ref{eq:ModeEqInEta} is therefore rather general and will form the basis of our analysis.

In a few cases equation \ref{eq:ModeEqInEta} can be solved exactly, but in general one can consider approximations in the super-Hubble and sub-Hubble regimes in which the wavelength of a mode is respectively much larger and much smaller than the Hubble radius, or equivalently $k\ll aH$ and $k\gg aH$ respectively. In the former case one can neglect the term in $k^2$ in equation \ref{eq:ModeEqInEta}, so that the equation has the general approximate solution $y_k=c_1a+c_2a\int d\eta\,a^{-2}$. The first expression is static, while the second is dynamic and decays fast, so that only the static terms remains relevant. In other words, the modes are `frozen'. For large values of $k$ on the other hand the equation is approximately that of a harmonic oscillator and such modes are evolving. During inflation, of course, a large range of modes pass from the sub-Hubble to the super-Hubble regime. 

Equation \ref{eq:ModeEqInEta} may be transformed to York time, giving
\begin{align} 0 &= y^{\prime\prime}_k+\left(\frac{1}{T}+2\frac{a^\prime}{a}-\frac{a^{\prime\prime}}{a}\right)y_k^\prime \notag\\
		&\qquad\qquad +\left[4M_{Pl}^4\frac{k^2}{a^2T^2}\frac{a^{\prime2}}{a^2}-\frac{a^\prime}{a}\left(\frac{1}{T}+2\frac{a^\prime}{a}\right)\right]y_k, \label{eq:ModeEqInT}
\end{align}
where primes denote derivatives with respect to $T$ and we used the relative time lapse,
\begin{equation} \DERIV{\eta}{T} = -\frac{2M_{Pl}^2a^\prime}{Ta^2},\end{equation}
obtained via the relationship between $T$ and $H$ \citep[see ref.\ ][]{RoserValentini2014a}. One can study equation \ref{eq:ModeEqInT} for distinct eras by choosing an appropriate function $a(T)$. During inflation the scale factor is given by equation \ref{eq:aduringinflation}, so that equation \ref{eq:ModeEqInT} takes the form
\begin{equation} 0 = y_k^{\prime\prime}-\left(T^{-1}+\kappa T\right)y_k^\prime + \kappa^2\left[4M_{Pl}^4\cdot\frac{k^2}{a^2}-2T^2\right]y_k, \label{eq:ModeEqInTInflation}\end{equation}
with $\kappa$ given by expression \ref{eq:kappadef}. Defining
\begin{align}\beta(T)&\equiv-(T^{-1}+\kappa T) \label{eq:betadef}\\ 
	     \omega^2_k(a,T)&\equiv\kappa^2[4M_{Pl}^4k^2/a^2-2T^2],  
\end{align}
equation \ref{eq:ModeEqInTInflation} may be cast into the form of a damped-oscillator equation ,
\begin{equation} 0= y_k^{\prime\prime}+\beta(T)y_k^\prime + \omega_k^2(a,T) y_k\end{equation}
(noting that $\beta(T)>0$ since $T<0$ while the universe is expanding). The coefficients are time-dependent, so this oscillator-like analysis does not provide solutions for $y_k$ over an extended period. But it does give an indication of the qualitative behaviour of a mode at some particular instant during inflation. Furthermore, inflation lasts for only a short York-time interval $\Delta T$ as compared to $|T|$, so that the evolution of the coefficients is, in fact, negligible apart from the time dependence entering via the scale factor.

The sign of $\omega_k^2$ now determines the nature of the evolution. If $\omega_k^2>0$, then $\omega_k$ is a real frequency and the evolution corresponds to damped oscillations. If $\omega_k^2<0$, the frequency is imaginary, corresponding to decay and hence a freezing of the mode. Physically the $k$-dependent term contributing to $\omega_k^2$ may be expressed in a more illuminating manner in terms of the ratio $\tilde{k}$ of the actual wave number $k$ with the wave number $aH$ of the mode instantaneously crossing the Hubble radius,
\begin{equation} \tilde{k} = \frac{k}{aH},\qquad  aH = -\tfrac12 M_{Pl}^{-2} Ta,\end{equation}
so that,
\begin{equation} \omega_k^2(a,T)= \left(\tilde{k}^2-2\right)\,\kappa^2T^2. \end{equation}
Super-Hubble modes satisfy $\tilde{k}\ll 1$, so that $\omega_k^2<0$ and the modes do not oscillate. In the sub-Hubble regime, $\tilde{k}\gg1$, the frequency is real, $\omega_k^2>0$, and the modes oscillate provided they are not overdamped (which is not the case, as we will see).\footnote{The fact that physically the regimes are identified by $\tilde{k}\ll1$ and $\tilde{k}\gg1$ while mathematically the critical value is $\sqrt{2}$ rather than $1$ is of no significance provided we restrict analysis to modes sufficiently far into the sub-Hubble or super-Hubble regime.}

We may study the effect of the damping term $\beta(T)$ by identifying whether a particular mode is overdamped or underdamped.\footnote{The possibility of critical damping may be neglected since the coefficients are time dependent and therefore no mode is critically damped for more than an instant.} The relevant quantity to consider is
\begin{align} D_k(a,T) &\equiv \beta^2(T)-4\omega_k^2(a,T) \notag\\
		       &= -16\kappa^2M_{Pl}^4\frac{k^2}{a_0^2}e^{\kappa T^2} + 9\kappa^2T^2+2\kappa +T^{-2},
\end{align}
where all terms are strictly positive except the first, which is strictly negative. Underdamping and overdamping correspond to $D_k<0$ and $D_k>0$ respectively. Since inflation occurs during a small York-time interval $\Delta T$, where $\Delta T/|T|\ll1$, one may approximate $T$ to be constant except for the $T$-dependence (entering via the scale factor) in the exponential, which decreases during inflation by a factor $e^{2N_e}$, $N_e$ being the number of e-folds given by equation \ref{eq:efolds}. Because the exponential term changes by a factor $N_e^2$ during the course of inflation, a set of modes with $k^2$ ranging over orders of magnitude given by this factor pass from being underdamped to being overdamped. This is, of course, conditional on the sign of $\omega_k^2$ not changing before the modes become overdamped. However, under reasonable assumptions they do. That is, the sign of $\omega_k^2$ changes before a mode becomes overdamped.

To see this last point, consider a mode with wave number $k$ such that at some initial York time $T_i$ near the beginning of inflation $\omega_k^2(a(T_i),T_i)>0$ and $D_k(a(T_i),T_i)<0$. The frequency changes from real to imaginary ($\omega_k^2=0$) roughly when crossing the Hubble radius, $\tilde{k}=\sqrt{2}\approx1$, that is, after an interval $\Delta T_{cross}$ given by
\begin{equation} e^{2\kappa T_i\Delta T_{cross}} = 2/\tilde{k}_i^2 ,\end{equation}
where $\tilde{k}_i$ refers to the value of $\tilde{k}$ corresponding to the mode $k$ at time $T_i$. Here only the leading-order change to the exponential has been considered, $\kappa T^2\approx \kappa T_i^2+2\kappa T_i\Delta T$, which is once again possible since $\Delta T/|T|\ll1$. Explicitly,
\begin{equation} \Delta T_{cross} = \left(2\kappa T_i\right)^{-1}\ln \left(2/k_i^2\right). \end{equation}
Meanwhile the mode passes from one damping regime to the other when $D_k=0$, or
\begin{equation} e^{2\kappa T_i\Delta T_{CD}}=\frac{9\kappa^2 T_i^2+2\kappa+T_i^{-2}}{16 M_{Pl}^4\kappa^2 k^2/a_i^2}, \label{eq:expdeltaTCD}\end{equation}
which may be further approximated by dropping the second and third term in the numerator if $\kappa^2T_i^2$ dominates. In section \ref{sec.timeline} we will provide numerical estimates showing that $\kappa^2T_i^2$ does indeed dominate, at least for some inflation models. One then has
\begin{equation} \Delta T_{CD} \approx \left(2\kappa T_i\right)^{-1}\ln \left(9/4\tilde{k}_i^2\right), \end{equation}
which is always strictly greater than $\Delta T_{cross}$. Thus the frequency will change from real to imaginary before the transition from underdamping to overdamping.

While it may be possible to construct conditions such that the above approximations are not valid and for which this result would not hold, the result does hold under conditions corresponding to inflation as it is presumed to occur in our actual universe. Hence the freezing-out process occurs, mathematically speaking, as a result of the change of sign of $\omega_k^2$ rather than a transition to overdamping.


\section{Re-entry in the radiation-dominated era} \label{sec.raddomuniverse}

According to our best understanding of cosmic history inflation was followed by a period of `reheating' during which the inflaton field decayed into the known particles of the standard model, although details of this process are uncertain. The energy density of the universe became subsequently dominated by radiation and during this period the scale factor behaved as $a\sim t^\frac12$. During this time the Hubble radius grew again as $\sim t$, allowing modes to `re-enter' and evolve after having previously been frozen. 

During the radiation-dominated era York time relates\footnote{Any explicit relationship between $t$ and $T$ can be modified by translation in $t$ since the cosmological equations in $t$ are time-translation invariant. However, the York-time theory is not $T$-translation invariant since $T$ has a physical meaning and the equations are explicitly $T$-dependent. The relationship given here therefore depends on the appropriate choice of time origin, namely that $t=0$ corresponds to $T=-\infty$.} to cosmological time as $T=-M_{Pl}^2/t$, so that the scale factor behaves as
\begin{equation} a(T) = a_{Pl}\frac{M_{Pl}}{|T|^\frac12}, \end{equation}
where $a_{Pl}$ is a proportionality constant formally referring to the scale factor at Planck York-time, $T=-M_{Pl}^2$.


The physical Hubble radius $H^{-1}$ is given via equation \ref{eq.THincosm}, 
which may be compared with the evolution of the physical wavelength of a mode,
\begin{equation} \lambda_{phys}(k) = a \lambda(k) = a_0\left(\frac{t}{t_0}\right)^\frac12\lambda = a_0\left(\frac{|T_0|}{|T|}\right)^\frac12\lambda,\end{equation}
where $\lambda$ is the comoving wavelength and $a_0$, $t_0$ and $T_0$ refer to the values of the corresponding quantities at some reference time (usually today). The Hubble radius grows faster than physical wavelengths and modes do indeed enter the Hubble radius during the radiation-dominated era.

The equation of motion for this period takes the form
\begin{equation} 0=y_k^{\prime\prime} + \frac{1}{2T} y_k^\prime + \frac{M_{Pl}^2k^2}{a_{Pl}^2|T|^3}y_k.\end{equation}
This can be solved exactly,
\begin{align} y_k(T)&=c_1\cdot\sqrt{|T|}e^{-i\frac{2M_{Pl}|k|}{a_{Pl}\sqrt{|T|}}}	\notag\\
		    &\qquad +c_2\cdot i\sqrt{|T|}\frac{a_{Pl}}{4M_{Pl}|k|}e^{i\frac{2M_{Pl}|k|}{a_{Pl}\sqrt{|T|}}}, \label{eq:yksolutionsradiation}
\end{align}
where $c_1$ and $c_2$ are arbitrary constants. The pre-factor of $\sqrt{T}$ results from `absorbing' a factor of $a$ into the variable $y$. The actual dynamical evolution is in the phases. A mode becomes `unfrozen' when its phase changes sufficiently rapidly. With $d/dT[\arg y_k]\sim|k||T|^{-\frac32}$, once unfrozen the speed of evolution only increases as $T\rightarrow0$ from below. Contrast this with the description in conformal time $\eta$. During the radiation-dominated era it follows from $a=a_0\sqrt{t/t_0}$ that $T=-M_{Pl}^2 t^{-1}$ and $dt\sqrt{t_0/t} = a_0 d\eta$ (so that $\eta=2\sqrt{tt_0}/a_0$ up to an additive constant). Hence the solutions \ref{eq:yksolutionsradiation} in terms of conformal time are
\begin{align} y_k(\eta) &= \bar{c}_1 \cdot \eta^{-1}e^{-i\frac{|k|}{a_{Pl}}\eta} \notag\\ 
			&\qquad +\bar{c}_2\cdot i\eta^{-1}\frac{a_{Pl}}{4M_{Pl}|k|} e^{i\frac{|k|}{a_{Pl}}\eta}, \end{align}
where $\bar{c}_1$ and $\bar{c}_2$ are arbitrary coefficients (related to $c_1$ and $c_2$ respectively by a constant factor). We see that the phase of each contribution is linear in $\eta$, so that the conformal `speed' of evolution $\deriv{\eta}(\arg y_k)$ is constant.


If the radiation-dominated period is taken to have begun at some point in time $\eta_r=2\sqrt{t_rt_0}/a_0=2M_{Pl}^2(a_0\sqrt{|T_r||T_0|})^{-1}$ when the scale factor had a value $a_r$, then the condition $\tilde{k}\ll1$ for a mode to be super-Hubble may be expressed as
\begin{equation} |k|\gg|T|^\frac12\cdot\pi a_r|T_r|^\frac12/M_{Pl}^2 = |T|^\frac12\cdot\text{const.}\end{equation}
When the inequality becomes an approximate equality,
\begin{equation} |k|\approx |T|^\frac12\cdot\pi a_r|T_r|^\frac12/M_{Pl}^2,\label{eq.modesstartevolving}\end{equation}
one expects the mode to evolve and indeed this is consistent with the solutions \ref{eq:yksolutionsradiation}.

Once a mode $k $ has become unfrozen ($\arg(y_k)\not\approx 0$), what are the time scales $\Delta \eta_k$ and $\Delta T_k$ for its evolution? In terms of conformal time $\eta$ the requirement for evolution to occur is 
\begin{equation} |k|\Delta\eta \approx a_0\eta_r, \end{equation}
while in terms of York time the condition \ref{eq.modesstartevolving} becomes $\Delta \sqrt{T} \approx\frac{|k|M_{Pl}^2}{a_r\sqrt{|T_r|}}$, or
\begin{equation} \Delta T_k \approx |T|\left( 1-\frac{|k|^2}{\left(|k|+a_rM_{Pl}^{-2}|T_r|^\frac12|T|^\frac12\right)^2}\right).\end{equation}
In the short-wavelength limit ($k\rightarrow\infty$), the York-time scale becomes small, $\Delta T_k\rightarrow0$ and so evolution is fast, as expected. Similarly, for modes in the super-Hubble regime ($k\rightarrow0$), which have not re-entered, one finds $\Delta T_k\rightarrow|T|$, that is, the York-time scale approaches the full duration of the remaining history of the universe.


\section{The cosmological timeline} \label{sec.timeline}

\begin{figure*}
  \includegraphics[width=\textwidth]{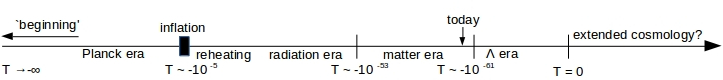}
  \caption{The cosmological timeline in terms of York time. The entire known cosmological history takes place within $10^{-5}$ of the end of the universe. If there is a cosmological constant then cosmological history ends just short of $T=0$.}
\end{figure*}

In 1969 Misner proposed that the quantity $-\ln a$ --- or (more or less equivalently) the logarithm of the temperature in the homogeneous approximation --- would provide a time parameter with which to give an account of cosmic history that is more adequate than the conventional cosmological time $t$ \citep{Misner1969c}. The new parameter would avoid the absurdly small numbers needed to describe early epochs,\footnote{Changing units to, say, Planck units leads to absurdly large numbers for later eras instead.} which in the language of $t$ took place within the first $10^{-33}$ (inflation) or even $10^{-43}$ (Planck era) of a second. Misner wrote: `The universe is meaningfully infinitely old, since infinitely many things have happened since the beginning.' While the argument is largely aesthetic, it shows that in a sense $t$ is a highly unnatural choice of time parameter when discussing the very early universe.

In terms of York time it is the early history that takes up the longest period $\Delta T$. In fact, just as in the case of Misner's parameter $-\ln a$, the `beginning' lies in the infinite past and unlike in the conventional description in terms of $t$ there is no notion of `before the Big Bang'. The Planck era stretches from $T=-\infty$ to just before the onset of inflation. Recent data \citep{Planck2015_ConstraintsOnInflation} leads to an upper bound on the Hubble parameter during inflation of $3.6\times10^{-5}M_{Pl}$, so that in reduced Planck units (where $M_{Pl}=1$) this number also gives a rough estimate of $T$. 

For the example considered in section \ref{sec.inflation} with the inflaton potential $V(\phi)=m^2\phi^2$, this empirical estimate for $H$ during inflation can be used to find an inequality (using equation \ref{eq:t0SRC}) for the kinetic-energy correction \ref{eq:KEcontribution},
\begin{equation} K.E. \ll 10^{-9}, \end{equation}
from which in turn one can derive an inequality for the constant $\kappa$,
\begin{equation} \kappa \gg 10^8.\end{equation}
This implies that the two expressions in the damping term $\beta(T)$ (equation \ref{eq:betadef}) are of roughly equal magnitude, while the term $\kappa^2T_i^2$ does indeed dominate in equation \ref{eq:expdeltaTCD}.

Following inflation and reheating, the energy density then becomes dominated by radiation until $t\sim10^4$ years. At this point matter becomes dominant and the Hubble parameter is of order $10^{-12}$s$^{-1}$, or $10^{-53}$ in reduced Planck units, which is already within the final moments of the universe when considered in terms of York time. Evidence suggests that today, at $T\sim H\sim 10^{-61}$, we are nearing a possibly final era dominated by a cosmological constant.

The York time description clearly does not do away with absurdly large order-of-magnitude differences in the way Misner might have envisioned. Rather, periods of large $T$-duration are of short $t$-duration and vice versa. The infinite `York age' of the universe might however be considered an aesthetic advantage.

One might speculate whether York time really `ends' at $T=0$ or whether cosmic history continues and our description ought to be extended. Indeed, if $T$ does have fundamental physical significance, then it is unwarranted to conclude that the universe would end then simply because the usual parameter $t$ ceases to describe such speculative future eras, although problems may arise if physical quantities are not well behaved during the transition to $T>0$. The existence of a cosmological constant also raises questions. In such a case the conventionally considered timeline never reaches $T=0$ but stops slightly short since $H(t)\rightarrow\Lambda$ (up to constants) as $t\rightarrow\infty$ rather than approaching zero. 


\section{Conclusion}\label{sec.conclusion}

Despite the prima facie difficulty---no York time passes during a true de~Sitter phase---we can account for inflation in the York time description by considering the next-order approximation, in which $H$ is slowly varying even during the inflationary period (although the York-time interval over which inflation occurs is then small). 

During inflation perturbations spanning a large range of wave numbers cross the Hubble radius from the sub-Hubble to the super-Hubble regimes. In the York-time description this manifests mathematically in that their frequency of oscillation becomes imaginary. The damping effect is found to be of only secondary importance with regards to understanding this process. It remains to be seen how these insights might contribute to an understanding of York-time cosmological perturbation theory when derived from first principles, a project begun in ref.\ \citep{Roser2015b}.

If York time is taken seriously as a candidate for a physically fundamental notion of time, for which, we argued, there are a number of theoretical reasons, then the timeline of cosmic history looks very different from the conventional description in terms of $t$. The Planck era ranges from $T=-\infty$ to the onset of inflation at $T\sim10^{-5}$ and the entirety of what is understood about the evolution of the universe lies between this time and $T=0$ (or even some instant a finite duration prior to $T=0$ in the case of a universe approaching another de~Sitter phase due to the effects of a cosmological constant). 

In the absence of a theory of quantum gravity it is considered an open question whether the notion of time makes sense for scales of the order of a Planck time and below --- in particular whether it makes sense to speak of a time before $t=10^{-43}$s. A theory based on a fundamental time parameter can answer this in the affirmative, where one may envision a quantised notion of space\footnote{At least the `shape' degrees of freedom are quantised. The local notion of scale, that is, the volume element takes the role of Hamiltonian density in the reduced-Hamiltonian formalism rather than that of a canonical variable and is therefore treated differently when quantising. At least for a minisuperspace model `volume' actually shows classical-like behaviour \citep{RoserValentini2014a}.} while $T$ continues to play a role as a continuous parameter. Ultimately the viability of such an approach must be decided empirically if possible. In the meantime, it is important to understand otherwise well-known phenomena in this framework in order to develop a physical intuition for the York-time description of cosmology. With the present paper we hope to contribute to this goal.



\setlength{\bibsep}{0pt plus 0.3ex}
\small
\bibliographystyle{unsrtnat}	
\bibliography{../../Bibloi}	

\begin{thebibliography}{33}
\providecommand{\natexlab}[1]{#1}
\providecommand{\url}[1]{\texttt{#1}}
\expandafter\ifx\csname urlstyle\endcsname\relax
  \providecommand{\doi}[1]{doi: #1}\else
  \providecommand{\doi}{doi: \begingroup \urlstyle{rm}\Url}\fi

\bibitem[Isham(1992)]{Isham1992}
C.\ Isham.
\newblock {Canonical quantum gravity and the problem of time}.
\newblock 1992.
\newblock Report No.: Imperial/TP/91-92/25, gr-qc/9210011.

\bibitem[Kucha\v{r}(2011)]{Kuchar2011}
K.~Kucha\v{r}.
\newblock {Time and interpretations of quantum gravity}.
\newblock \emph{International Journal of Modern Physics D}, 20:\penalty0 3--86,
  2011.

\bibitem[Anderson(2012)]{Anderson2012}
E.~Anderson.
\newblock {The problem of time in quantum gravity}.
\newblock In V.~R. Frignanni, editor, \emph{{Classical and Quantum Gravity:
  Theory, Analysis and Applications}}. Nova (New York), 2012.
\newblock gr-qc: 1206.2403.

\bibitem[Liberati(2013)]{Liberati2013}
S.~Liberati.
\newblock {Tests of Lorentz invariance: a 2013 update}.
\newblock \emph{Classical and Quantum Gravity}, 30:\penalty0 133001, 2013.
\newblock gr-qc: 1304.5795.

\bibitem[Valentini(2008)]{Valentini2008inCraigSmith}
A.~Valentini.
\newblock {Hidden variables and the large-scale structure of spacetime}.
\newblock In W.L. Craig and Q.~Smith, editors, \emph{Einstein, Relativity and
  Absolute Simultaneity}. Routledge, 2008.

\bibitem[Afshordi(2015)]{Afshordi2015}
N.~Afshordi.
\newblock {Why is high-energy physics Lorentz invariant?}
\newblock 2015.
\newblock hep-th: 1511.07879.

\bibitem[Colladay and Kosteleck\'{y}(1997)]{ColladayKostelecky1997}
D.~Colladay and A.~Kosteleck\'{y}.
\newblock {CPT violation and the Standard Model}.
\newblock \emph{Physical Review D}, 55:\penalty0 6760, 1997.
\newblock hep-th: 9703464.

\bibitem[Kosteleck\'{y}(2004)]{Kostelecky2004}
Alan Kosteleck\'{y}.
\newblock {Lorentz violation and gravity}.
\newblock In \emph{{Third Meeting on CPT and Lorentz Symmetry}}, 2004.
\newblock hep-ph: 0412406.

\bibitem[Amelino-Camelia(2004)]{Amelino-Camelia2004}
G.~Amelino-Camelia.
\newblock {Phenomenology of Planck-scale Lorentz-symmetry test theories}.
\newblock \emph{New Journal of Physics}, 6:\penalty0 188, 2004.

\bibitem[Liberati and Maccione(2009)]{LiberatiMaccione2009}
S.~Liberati and L.~Maccione.
\newblock {Lorentz violation: motivation and new constraints}.
\newblock \emph{Annual Review of Nuclear and Particle Science}, 59:\penalty0
  245--267, 2009.
\newblock astro-ph: 0906.0681.

\bibitem[Ho\v{r}ava(2009)]{Horava2009}
P.~Ho\v{r}ava.
\newblock {Quantum gravity at a Lifshitz point}.
\newblock \emph{Physical Review D}, 79:\penalty0 084008, 2009.
\newblock https://arxiv.org/abs/0901.3775.

\bibitem[Visser(2011)]{Visser2011}
M.~Visser.
\newblock {Status of Ho\v{r}ava gravity: A personal perspective}.
\newblock 2011.
\newblock hep-th: 1103.5587.

\bibitem[Afshordi(2009)]{Afshordi2009}
N.~Afshordi.
\newblock {Cuscuton and low energy limit of Horava-Lifshitz gravity}.
\newblock \emph{Physical Review D}, 80:\penalty0 081502, 2009.
\newblock hep-th: 0907.5201.

\bibitem[Marsden and Tipler(1980)]{MarsdenTipler1980}
J.E. Marsden and F.J. Tipler.
\newblock {Maximal hypersurfaces and foliations of constant mean curvature in
  general relativity}.
\newblock \emph{Physics Reports}, 66:\penalty0 109, 1980.

\bibitem[Choquet-Bruhat and York(1980)]{ChoquetBruhatYork1980}
Y.~Choquet-Bruhat and J.~York.
\newblock {The Cauchy problem}.
\newblock In A.~Held, editor, \emph{General Relativity and Gravitation I}.
  Plenum, 1980.

\bibitem[Arnowitt et~al.(1962)Arnowitt, Deser, and Misner]{ADM1962}
R.~Arnowitt, S.~Deser, and C.~W. Misner.
\newblock The dynamics of general relativity.
\newblock In L.~Witten, editor, \emph{Gravitation: an introduction to current
  research}. Wiley, 1962.

\bibitem[Misner et~al.(1973)Misner, Thorne, and
  Wheeler]{MisnerThorneWheeler1973}
C.~Misner, K.~Thorne, and J.~Wheeler.
\newblock \emph{Gravitation}.
\newblock W.H. Freeman, 1973.

\bibitem[Roser and Valentini(2014)]{RoserValentini2014a}
P.~Roser and A.~Valentini.
\newblock {Classical and quantum cosmology with York time}.
\newblock \emph{Classical and Quantum Gravity}, 31:\penalty0 245001, 2014.
\newblock gr-qc: 1406.2036.

\bibitem[Roser(2016{\natexlab{a}})]{Roser2015a}
P.~Roser.
\newblock {Quantum mechanics on York slices}.
\newblock \emph{Classical and Quantum Gravity}, 33:\penalty0 065001,
  2016{\natexlab{a}}.
\newblock qr-qc: 1507.01556.

\bibitem[Roser(2015)]{Roser2015b}
P.~Roser.
\newblock {Cosmological perturbation theory with York time}.
\newblock \emph{arxiv gr-qc: 1511.03320}, 2015.
\newblock gr-qc: 1511.03320.

\bibitem[York(1972)]{York1972}
J.~York.
\newblock Role of conformal three-geometry in the dynamics of gravitation.
\newblock \emph{Physical Review Letters}, 28:\penalty0 1082--1085, 1972.

\bibitem[Gomes et~al.(2011)Gomes, Gryb, and Koslowski]{GomesGrybKoslowski2011}
H.~Gomes, S.~Gryb, and T.~Koslowski.
\newblock Einstein gravity as a 3d conformally invariant theory.
\newblock \emph{Classical and Quantum Gravity}, 28:\penalty0 045004, 2011.
\newblock gr-qc: 1010.2481.

\bibitem[Barbour et~al.(2014)Barbour, Koslowski, and
  Mercati]{BarbourKoslowskiMercati2013ProbOfTime}
J.~Barbour, T.~Koslowski, and F.~Mercati.
\newblock {The solution to the problem of time in shape dynamics}.
\newblock \emph{Classical and Quantum Gravity}, 31:\penalty0 155001, 2014.
\newblock gr-qc: 1302.6264.

\bibitem[Mercati(2014)]{Mercati2014}
F.~Mercati.
\newblock {A shape dynamics tutorial}.
\newblock 2014.
\newblock gr-qc: 1409.0105v1.

\bibitem[Qadir and Wheeler(1985)]{QadirWheeler1985}
A.~Qadir and J.A. Wheeler.
\newblock In \emph{{From $SU(3)$ to Gravity}}. Cambridge University Press,
  1985.

\bibitem[Roser(2016{\natexlab{b}})]{Roser2015CosmExtension}
P.~Roser.
\newblock {An extension of cosmological dynamics with York time}.
\newblock \emph{General Relativity and Gravitation}, 48\penalty0 (4):\penalty0
  1--15, 2016{\natexlab{b}}.
\newblock gr-qc: 1407.4005.

\bibitem[Valentini(1996)]{Valentini1996}
A.~Valentini.
\newblock {Pilot-wave theory of fields, gravitation and cosmology}.
\newblock In J.~T. Cushing, A.~Fine, and S.~Goldstein, editors, \emph{Bohmian
  mechanics and quantum theory: an appraisal}. Kluwer, 1996.

\bibitem[{Planck Collaboration}(2015)]{Planck2015_Overview}
{Planck Collaboration}.
\newblock {Planck 2015 results. I. Overview of products and scientific
  results}.
\newblock 2015.
\newblock astro-ph: 1502.01582v1.

\bibitem[Guth(1981)]{Guth1981}
A.~Guth.
\newblock {Inflationary universe: a possible solution to the horizon and
  flatness problems}.
\newblock \emph{Physical Review D}, 23:\penalty0 347, 1981.

\bibitem[Brandenberger and Peter(2016)]{BrandenbergerPeter2016}
R.H. Brandenberger and P.~Peter.
\newblock {Bouncing Cosmologies: Progress and Problems}.
\newblock 2016.
\newblock hep-th: 1603.05834.

\bibitem[Muhkanov et~al.(1992)Muhkanov, Feldman, and
  Brandenberger]{MukhanovFeldmanBrandenberger1992}
V.F. Muhkanov, H.A. Feldman, and R.H. Brandenberger.
\newblock {Theory of cosmological perturbations}.
\newblock \emph{Physics Reports}, 215:\penalty0 203, 1992.

\bibitem[Misner(1969)]{Misner1969c}
C.~Misner.
\newblock {Absolute zero of time}.
\newblock \emph{Physical Review}, 186:\penalty0 1328--1333, 1969.

\bibitem[Collaboration(2015)]{Planck2015_ConstraintsOnInflation}
Planck Collaboration.
\newblock {Planck 2015 results. XX. Constraints on Inflation}.
\newblock 2015.
\newblock astro-ph: 1502.02114.

\end{thebibliography}


\end{multicols}

\end{document}